\documentclass[onecolumn,pra,preprint]{revtex4}
\usepackage{epsf}
\usepackage{amsmath,amsthm,amssymb}
\usepackage{color}
\usepackage{dcolumn}
\usepackage{bm}
\usepackage{graphicx,epsfig}
\usepackage[dvipsnames]{xcolor}

\graphicspath{{fig/}}
\everymath{\displaystyle}

\begin{document}           

\title{Relativistic quantum dynamics of twisted electron beams in arbitrary electric and magnetic fields}

\author{Alexander J. Silenko $^{1,2,3}$} \email{alsilenko@mail.ru}
\author{Pengming Zhang$^{1,4}$}
\email{zhpm@impcas.ac.cn}
\author{Liping Zou$^{1,4}$}
\email{zoulp@impcas.ac.cn}

\affiliation{$^1$Institute of Modern Physics, Chinese Academy of Sciences, Lanzhou 730000, China}
\affiliation{$^2$Bogoliubov Laboratory of Theoretical Physics, Joint Institute for Nuclear Research, Dubna 141980, Russia}
\affiliation{$^3$Research Institute for Nuclear Problems, Belarusian State University, Minsk 220030, Belarus}
\affiliation{$^4$University of Chinese Academy of Sciences, Yuquanlu 19A, Beijing 100049, China}

\date{\today}

\begin{abstract}
Relativistic quantum dynamics of twisted (vortex) Dirac particles in arbitrary electric and magnetic fields is constructed for the first time.
This allows to change the controversial contemporary situation when the nonrelativistic approximation is used for relativistic twisted electrons.
The relativistic Hamiltonian and equations of motion in the Foldy-Wouthuysen representation are derived.
A critical experiment for a verification of the results obtained is proposed. The new important
effect of a radiative orbital polarization of a twisted electron beam in a magnetic field resulting
in a nonzero average projection of the intrinsic orbital angular momentum on the field direction is
predicted.
\end{abstract}

\maketitle

The discovery of twisted (vortex) electron beams carrying an intrinsic orbital
angular momentum (OAM) \cite{UTV} which existence has been predicted in Ref. \cite{Bliokh2007}
opens new possibilities in the electron microscopy and in investigations of magnetic phenomena
(see Refs. \cite{BliokhSOI,Lloyd,LloydPhysRevLett2012,Rusz,Edstrom,imaging,Observation,OriginDemonstration}
and references therein). Twisted electron beams with large intrinsic OAMs (up to 1000$\hbar$)
have been recently obtained \cite{VGRILLO}. At present, much attention is also devoted to the
interactions of such beams with atoms, nuclei and a laser field \cite{electpart}.
The dynamics of the intrinsic OAM in external magnetic and electric fields has been studied in Refs. \cite{Bliokh2007,Bliokhmagnetic,magnetic,Kruining,Rajabi,classicalmagnetic,experimentmagnetic}.
However, we note in advance that our result is different from the equation of motion of the
intrinsic OAM in an electric field previously found in Ref. \cite{Bliokh2007}.
The general
description of the relativistic dynamics of an intrinsic OAM in arbitrary electric and magnetic fields in the framework of classical physics has
been made in Ref. \cite{Manipulating}. In particular, the correction of the equation of
motion of the intrinsic OAM in the electric field previously obtained in Ref. \cite{BliokhSOI}
has been fulfilled in this work. The methods of the manipulation of beams developed in
Ref. \cite{Manipulating} use not only the magnetic field but also the electric one.
The great importance of twisted electrons requires detailed investigations of fundamental quantum-mechanical
properties of twisted electron beams. Such investigations have been carried out in many
publications (see Refs. \cite{BliokhSOI,Lloyd} and references therein). However, only the nonrelativistic approximation has been formerly used. The correct
quantum dynamics of a twisted electron in an \emph{electric} field has not been elaborated at all. Therefore, some
fundamental properties need a further inquiry. The quantum-mechanical approach used differs the present work from our recent study \cite{Manipulating} where the classical approach has been applied.

In the present work, the system of units $\hbar=1,~c=1$ is used. We include $\hbar$ and
$c$ explicitly when this inclusion clarifies the problem. The square and curly brackets,
$[\dots,\dots]$ and $\{\dots,\dots\}$, denote commutators and anticommutators, respectively.

Since vortex electrons are relativistic quantum objects admitting also a semiclassical
description, a construction of a \emph{relativistic} Schr\"{o}dinger-like dynamics of
such particles is necessary. It is important that one mainly observes a motion of charged
centroids and it is instructive to mirror this circumstance in an appropriate quantum description.
We solve this problem in the present work. We should also notice the previous quantum-mechanical
analysis mirrored in the reviews \cite{BliokhSOI,Lloyd}.

 A twisted electron is a single pointlike particle. However, its wave function is a
 superposition of states with different momentum directions. While the twisted electron
 in vacuum has a nonzero intrinsic OAM and a nonzero component of the momentum in the
 plane orthogonal to the direction of its resulting motion, it can be described by the
 Dirac equation for a free particle.
Quantum mechanics (QM) of the twisted electron in external electromagnetic fields is
also governed by the usual Dirac equation.
We can disregard the anomalous magnetic moment of the electron because its $g$ factor
is close to 2. The Schr\"{o}dinger form of the relativistic QM is provided by the
relativistic Foldy-Wouthuysen (FW) transformation. The relativistic extensions of the FW method \cite{FW} have been first proposed in Refs. \cite{Case,B}. There are many other methods of the relativistic FW transformation (see Refs. \cite{SuttorpdeGroot,TMP,JMP,BliokhK,Goss,local,relativisticFW,ChiouChen}
and references therein). The results obtained by different methods agree because of the uniqueness of the FW representation proven in Ref. \cite{E}.
In all previous studies of twisted particles in external fields, only the nonrelativistic FW transformation has been fulfilled (see the reviews \cite{BliokhSOI,Lloyd}). The relativistic FW transformation has been formerly used only for free twisted particles \cite{BarnettPRL,BliokhPRA2017}. Thus,
the nonrelativistic approximation has been used for a description of relativistic objects (the kinetic energy of twisted electrons, $200-300$ keV, is comparable with their rest energy, 511 keV).
As a result, the present state of the theory of twisted electron beams is controversial.

The \emph{exact} relativistic Hamiltonian in the FW representation
(the FW Hamiltonian) for a Dirac particle in a magnetic field has been first obtained in Ref. \cite{Case} and its validity has been confirmed in other works \cite{JMP,Energy1,Energy3}. It is given by
\begin{equation}
\begin{array}{c}
{\cal H}_{FW}=\beta\sqrt{m^2+\bm{\pi}^2-e\bm\Sigma\cdot\bm B},
\end{array}
\label{eq33new}
\end{equation}
where $\bm{\pi}=\bm{p}-e\bm A$ is the kinetic momentum, $\bm B=\nabla\times \bm A$
is the magnetic induction, and $\beta$ and $\bm\Sigma$ are the Dirac matrices. This Hamiltonian is valid for a twisted and a untwisted particle.
The spin angular momentum operator is equal to $\bm s=\hbar\bm\Sigma/2$. The magnetic field is, in general, nonuniform but time-independent.
Unlike all previous investigations, our quantum description of a twisted Dirac particle is fully relativistic
and correctly defines both electric and magnetic interactions. Our precedent
study \cite{Manipulating} has shown the importance of the electric field
for the manipulation of twisted electron beams.

The FW Hamiltonian acts on the bispinor wave function $\Psi_{FW}=
\left(\begin{array}{c} \phi \\ 0 \end{array}\right)$. Since both the nonrelativistic and the relativistic FW Hamiltonians commute with the operators $\bm{\pi}^2$ and $s_z$, their eigenfunctions coincide. In
the \emph{uniform} magnetic field $\bm B=B\bm e_z$, they have
the form of non-diffracting Laguerre-Gauss beams \cite{Bliokhmagnetic,LL3,Greenshields}. 
It is convenient to present the exact energy spectrum obtained by different methods \cite{Case,Energy1,Energy3,Energy2a,Energy2,OConnell,CanutoChiu} as follows:
\begin{equation}
\begin{array}{c}
E=\sqrt{m^2+(2n+1+|l_z|+l_z+2s_z)|e|B},
\end{array}
\label{eqOAM}
\end{equation} where $n=0,1,2,\dots$ is the radial quantum number and $\bm l=\bm r\times\bm p=\bm L+\bm L^{(e)}$ is the total
OAM operator being the sum of the intrinsic ($\bm L$) and extrinsic ($\bm L^{(e)}$) OAMs. The same energy spectrum but for the squared energy operator has been obtained in Refs. \cite{Kruining,Rajabi}. We consider the conventional canonical OAMs. The difference between the dynamics of the canonical and kinetic (mechanical) OAMs has been investigated in Ref. \cite{Barnett}.
All energy levels of twisted particles in the uniform magnetic field
belong to the Landau levels. The relativistic approach (unlike the nonrelativistic one) demonstrates that the Landau levels are not equidistant for any field strength. This property has not been mentioned in Refs. \cite{Kruining,Rajabi,Case,Energy1,Energy3,Energy2a,Energy2,OConnell,CanutoChiu} while it has been noted in Ref. \cite{Energy2} that the energy spectrum becomes quasicontinuous when the quantum number $n$ becomes very large. A nonzero intrinsic OAM increases a degeneracy
multiplicity of the Landau levels (cf. Refs. \cite{BliokhSOI,Bliokhmagnetic}).

As a rule, we can use the weak-field approximation and can suppose that the
de Broglie wavelength, $\hbar/p$, is much smaller than the characteristic
size of the nonuniformity region of the external field. In this case, the
Hamiltonian (\ref{eq33new}) takes the form
\begin{eqnarray}
{\cal H}_{FW}&=&\beta\sqrt{m^2+\bm{\pi}^2}
-\frac e4\left\{\frac{1}{\epsilon'},\bm\Pi\cdot\bm B\right\}
\nonumber \\
&=&\beta\epsilon'-\beta\frac {e}{4}
\left[\frac{1}{\epsilon'}(\bm l+\bm\Sigma)\cdot\bm B+\bm B\cdot(\bm l+\bm\Sigma)\frac{1}{\epsilon'}\right],
\nonumber \\
\epsilon'&=&\sqrt{m^2+\bm{p}^2},
\label{eqHmOAM}
\end{eqnarray}
where $[\epsilon',\bm l]=0$.
Equation (\ref{eqHmOAM})
agrees with the result obtained in Ref. \cite{Greenshields}. The small term $m\omega_L^2r^2/2$
presented in Eq. (2) of Ref. \cite{Greenshields} expresses the rotational energy of the
centroid \emph{in the magnetic field}. This term originates from $e^2\bm A^2/(2m)$ and
is omitted in the present study because it is proportional to $B^2$.

It is necessary to take into account that a twisted electron is a charged centroid
\cite{Bliokh2007,BliokhSOI}. To describe observable quantum-mechanical effects, we need to present the
Hamiltonian in terms of the centroid parameters. The centroid as a whole is characterized
by the center-of-charge radius vector $\bm R$ and by the kinetic momentum
$\bm{\pi}'=\bm P-e\bm A(\bm R)$, where $\bm P=-i\hbar\partial/(\partial\bm R)$.
The intrinsic motion is defined by the kinetic momentum
$\bm\pi''=\pmb{\mathfrak{p}}-e[\bm A(\bm r)-\bm A(\bm R)]$. Here
$\pmb{\mathfrak{p}}=-i\hbar\partial/(\partial\pmb{\mathfrak{r}})$,
$\pmb{\mathfrak{r}}=\bm r-\bm R$, $\bm{\pi}'+\bm{\pi}''=\bm{\pi}$,
$\bm P+\pmb{\mathfrak{p}}=\bm p$. Since
\[
\bm A(\bm r)=\bm A(\bm R)+\frac12\bm B(\bm R)\times\pmb{\mathfrak{r}},
\]
the
operator $\bm{\pi}^2$ takes the form
\[
\bm{\pi}^2={\bm{\pi}'}^2+{\pmb{\mathfrak{p}}}^2
-\frac e2\left[\bm L\cdot\bm B(\bm R)+\bm B(\bm R)\cdot\bm L\right]
+\bm{\pi}'\cdot\bm{\pi}''+\bm{\pi}''\cdot\bm{\pi}'.
\]
After summing over partial waves with different momentum directions,
$<\bm{\pi}'\cdot\bm{\pi}''+\bm{\pi}''\cdot\bm{\pi}'>=0$. More precisely, the operator
$\bm{\pi}'\cdot\bm{\pi}''+\bm{\pi}''\cdot\bm{\pi}'$ has zero expectation values for
any eigenstates of the operator $\bm{\pi}^2$ and, therefore, it can be omitted.
It can be added that this summing can be performed for the squared Hamiltonian ${\cal H}_{FW}^2$.

The FW Hamiltonian summed over the partial waves \cite{BliokhSOI} takes the form
\begin{eqnarray}
{\cal H}_{FW}&=&\beta\epsilon 
-\beta\frac {e}{4}\left[\frac{1}{\epsilon}\bm\Lambda\cdot\bm B(\bm R)
+\bm B(\bm R)\cdot\bm\Lambda\frac{1}{\epsilon}\right],
\nonumber \\
\epsilon&=&\sqrt{m^2+{\bm{\pi}'}^2+{\pmb{\mathfrak{p}}}^2},
\qquad \bm\Lambda=\bm L+\bm\Sigma.
\label{eqHmsum}
\end{eqnarray}
The momentum and the intrinsic OAM can have different mutual orientations
in different Lorentz frames \cite{Manipulating,BliokhHallEffect}. Such a
geometry of twisted waves has been described in Ref. \cite{BliokhNori}.

The acceleration of the twisted electron in a uniform magnetic field does not
depend on $\bm\Lambda$. However, such a dependence takes place in a nonuniform
magnetic field due to the Stern-Gerlach-like force defined by the operator
\begin{equation}
\begin{array}{c}
\bm F_{SGl}=\beta\frac e4\left(\frac{1}{\epsilon}\nabla\bigl[\bm\Lambda\cdot\bm B(\bm R)\bigr]
+\nabla\bigl[\bm B(\bm R)\cdot\bm\Lambda\bigr]\frac{1}{\epsilon}\right),
\end{array}
\label{force}
\end{equation}
where $\nabla\equiv\partial/(\partial\bm R)$. This force is an analog of
the Stern-Gerlach one affecting the spin but it is much (approximately, $2L$ times) stronger.
The operators of the magnetic and electric dipole moments,
$\bm\mu$ and $\bm d$, are defined by
\begin{equation}
\begin{array}{c}
{\cal H}^{(int)}_{FW}=-\frac12\bigl[\bm\mu\cdot\bm B(\bm R)
+\bm B(\bm R)\cdot\bm\mu+\bm d\cdot\bm E(\bm R)+\bm E(\bm R)\cdot\bm d\bigr],
\end{array}
\label{mdmedmo}
\end{equation}
where ${\cal H}^{(int)}_{FW}$ is the interaction Hamiltonian. The operator
of the magnetic dipole moment of a moving centroid obtained from Eq. (\ref{eqHmsum})
is given by
\begin{equation}
\begin{array}{c}
\bm\mu=\beta\frac{e(\bm L+2\bm s)}{2\epsilon}.
\end{array}
\label{mdm}
\end{equation}
This equation agrees with Refs. \cite{Kruining,Manipulating,Barut,BliokhPRL2011}.
When the weak-field approximation is not used, it can be obtained that the
magnetic dipole moment of a twisted particle in a magnetic field is proportional
to the kinetic OAM $\bm r\times\bm\pi$ and is affected by the additional current.
This current is proportional to $-e\bm A$ and leads to a weak diamagnetic effect
\cite{Bliokh2007,BliokhSOI,Greenshields2015}.

To perform a general quantum description of a twisted Dirac particle in external
fields, we need to add terms dependent on the electric field. For a pointlike
Dirac particle, these terms have been obtained in Refs. \cite{relativisticFW,TMP,RPJ}.
If one disregards spin effects, one needs to add only the term $e\Phi(\bm r)$.
It has been proven in Ref. \cite{JINRLett12} that the passage to the classical
limit in the FW representation
reduces to a replacement of the operators in quantum-mechanical Hamiltonians
and equations of motion
with the corresponding classical quantities. The motion of the intrinsic OAM
causes the electric dipole moment $\bm d$ which interaction with the electric
field should be taken into account. Certainly, $\bm d^{(0)}=0$ in the rest
frame of the twisted electron. The results obtained in Refs. \cite{Manipulating,PhysScr}
show that in the classical limit
\begin{equation}
\begin{array}{c}
\bm d=\bm\beta\times\bm\mu,\qquad \bm\beta=\frac{\bm V}{c}\equiv \frac{\dot{\bm R}}{c}.
\end{array}
\label{eqHdOAM}
\end{equation}
The centroid velocity operator, $\bm V$, can be obtained from the FW Hamiltonian:
\[
\bm V=i[{\cal H}_{FW},\bm R]=\frac\beta2\left\{\frac{1}{\epsilon},\bm\pi'\right\}.
\]
Since we use the weak-field approximation, we neglect the correction to this
formula proportional to the electric field.
A comparison with Refs. \cite{relativisticFW,TMP,RPJ} allows us to obtain the
formula for the operator of the electric dipole moment:
\begin{equation}
\begin{array}{c}
\bm d=\beta\frac{e}{4}\left(\frac{1}{\epsilon}\bm\beta\times\bm L
-\bm L\times\bm\beta\frac{1}{\epsilon}\right)=\frac{e}{4}\left(\frac{1}{\epsilon^2}
\bm\pi'\times\bm L
-\bm L\times\bm\pi'\frac{1}{\epsilon^2}\right).
\end{array}
\label{edmoper}
\end{equation}

Thus, the general FW Hamiltonian for a relativistic twisted particle in electric
and magnetic fields is given by
\begin{equation}
\begin{array}{c}
{\cal H}_{FW}=\beta\epsilon+e\Phi 
-\beta\frac{e}{4}\biggl[\frac{1}{\epsilon}\bm L\cdot\bm B(\bm R)
+\bm B(\bm R)\cdot\bm L\frac{1}{\epsilon}\biggr] \\ 
+\frac{e}{4}\biggl\{\frac{1}{\epsilon^2}\bm L\cdot[\bm\pi'\times\bm E(\bm R)]
-[\bm E(\bm R)\times\bm\pi']\cdot\bm L\frac{1}{\epsilon^2}\biggr\}.
\end{array}
\label{eqHdsum}
\end{equation}
In this equation, spin effects are disregarded because they can be neglected on the
condition that $L\gg1$. The term $e\Phi$ does not include the interaction of the
intrinsic OAM with the electric field.

Equation (\ref{eqHdsum}) exhaustively describes the quantum dynamics of the
intrinsic OAM in the general case of a twisted Dirac particle in arbitrary
electric and magnetic fields. In particular, the equation of motion of the
intrinsic OAM has the form
\begin{eqnarray}
\frac{d\bm L}{dt}&=&i[{\cal H}_{FW},\bm L]=\frac12
\left(\bm\Omega\times\bm L-\bm L\times\bm\Omega\right),
\nonumber \\
\bm\Omega&=&-\beta\frac{e}{4}\left\{\frac{1}{\epsilon},\bm B(\bm R)\right\} 
+
\frac{e}{4}\left[\frac{1}{\epsilon^2}\bm\pi'\times\bm E(\bm R)-\bm E(\bm R)
\times\bm\pi'\frac{1}{\epsilon^2}\right].
\label{DynaOAM}
\end{eqnarray}
When $\bm E(\bm R)=0$, we obtain the relativistic quantum-mechanical equation
for the Larmor precession.
A comparison with results obtained in the Dirac representation shows strengths of the FW one. In Refs. \cite{Kruining,Rajabi}, a relativistic description of twisted electron beams in a uniform magnetic field has been given in the
former representation. However, the corresponding equation of motion of the OAM has not been obtained in these works.

Another important problem is the relativistic quantum dynamics of the kinetic momentum.
It is defined by the force operator:
\begin{eqnarray}
\bm F &=&\frac {d\bm\pi'}{dt}=\frac {\partial\bm\pi'}{\partial t}+i[{\cal H}_{FW},\bm\pi'] \nonumber \\
&=&
e\bm E(\bm R)+\beta\frac{e}{4}\Biggl\{\frac{1}{\epsilon},
\Bigl(\bm\pi'\times\bm B(\bm R)-\bm B(\bm R)\times\bm\pi'\Bigr)\Biggr\}+\bm F_{SGl}.
\label{forcetot}
\end{eqnarray}
Beam splitting in nonuniform electric and magnetic fields is conditioned
by the Stern-Gerlach-like force operator
\begin{eqnarray}
\bm F_{SGl}=&&\beta\frac {e}{4}\biggl\{\frac{1}{\epsilon}\nabla[\bm L\cdot\bm B(\bm R)]+\nabla[\bm B(\bm R)\cdot\bm L]\frac{1}{\epsilon}\biggr\} \nonumber \\
&&-\frac {e}{4}\biggl\{\frac{1}{\epsilon^2}\nabla(\bm L\cdot[\bm\pi'\times\bm E(\bm R)])-\nabla([\bm E(\bm R)\times\bm\pi']\cdot\bm L)\frac{1}{\epsilon^2}\biggr\}.
\label{forcegen}
\end{eqnarray}
The force is exerted to the center of charge of the centroid. Equations (\ref{force}) and  (\ref{forcegen}) demonstrate an importance of a general description which includes particle interactions with nonuniform fields.
Like Eq. (\ref{force}),
Eq. (\ref{forcegen}) presents the Stern-Gerlach-like force in terms of the centroid parameters.
It is important to mention that $\sqrt{m^2+{\pmb{\mathfrak{p}}}^2}$ is an effective mass of the twisted particle.

We can conclude that the quantum-mechanical equations obtained in the present work
agree with the corresponding classical results given in Ref. \cite{Manipulating}.

In the present work, we consider two important applications of the 
results obtained and describe new effects which experimental study allows one to ascertain
fundamental properties of twisted electron beams.

\textit{Rotation of the intrinsic OAM in crossed electric and magnetic fields.}--
Previously
fulfilled experiments used coherent superpositions of two twisted beams moving in a
longitudinal magnetic field ($z$ axis). Landau modes with equal amplitudes, the same
radial index $n$, and opposite projections of the intrinsic longitudinal OAMs undergo
the rotation with the Larmor frequency \cite{BliokhSOI,Bliokhmagnetic,Greenshields}.
This rotation is similar to the Faraday effect in optics \cite{Greenshields}.
We propose a similar and a simpler experiment in crossed electric and magnetic fields
satisfying the relation $\bm E=-\bm\beta\times\bm B$, where $\bm E\bot\bm B\bot\bm\beta$ and $\bm\beta$
is the normalized beam velocity. Such fields characterizing the Wien filter do not affect
a beam trajectory. We suppose the fields $\bm E$ and $\bm B$ to be uniform.

In the considered case, the classical limit of the relativistic equation for the angular
velocity of precession of the intrinsic OAM is given by
\begin{equation}
\begin{array}{c}
 \bm\Omega^{(W)}=-\frac{e(m^2+{\pmb{\mathfrak{p}}}^2)}{2\epsilon^3}\bm B.
\end{array}
\label{reLW}
\end{equation}
Since ${\pmb{\mathfrak{p}}}^2\ll m^2$, Eq. (\ref{reLW}) agrees with the corresponding
equation obtained in Ref. \cite{Manipulating}.

While the experiment proposed is similar to the above-mentioned experiment carried
out in Refs. \cite{Bliokhmagnetic,Greenshields}, it needs a simpler experimental setup.
It is necessary to use a single twisted electron beam possessing a standard orbital
polarization collinear to the beam momentum ($z$ axis). The direction of the magnetic
field $\bm B$ and the quasimagnetic one $\bm\beta\times\bm E$ is transversal ($x$ axis).
Therefore, the intrinsic OAM rotates in the $yz$ plane with the angular
frequency $\bm\Omega^{(W)}$ and reverses its direction with the angular
frequency $2\bm\Omega^{(W)}$. Since twisted electron beams are relativistic, a
quantitative verification of Eq. (\ref{reLW}) can be fulfilled. Thus, the experiment
proposed is a \emph{critical experiment} for a verification of the main equations
for the FW Hamiltonian and the intrinsic-OAM dynamics obtained in the present work.

\textit{Radiative orbital polarization of twisted electron beams in a magnetic field.}--
The
radiation from twisted electrons is one of their fundamental properties. It is caused by
their acceleration in external magnetic and electric fields and by the time dependence of
their magnetic moments. The latter effect conditions the magnetic dipole radiation
\cite{TransitionRadiation}. Its intensity is standardly much smaller than that of the
electric dipole radiation due to the particle acceleration. The magnetic dipole radiation
proportional to the intrinsic OAMs is important for the Cherenkov radiation and the
transition one \cite{TransitionRadiation}. Since the particle motion in a magnetic field
is accelerated, we can consider only the electric dipole radiation. For particles closed
in storage rings, it is conditioned by radiative transitions between Landau levels and is
called the synchrotron radiation. As a rule, magnetic focusing is used. We expect that a
storage and an acceleration of twisted electrons in cyclotrons (electron rings) will be
carried out in a nearest future. Evidently, these processes cannot vanish the intrinsic
OAM due to the conservation of angular momentum. A natural process of an orbital depolarization
is caused by a loss of the beam coherence and is not important for the considered problem.
The initial orbital polarization of twisted electrons is longitudinal ($z$ axis) while
their expected final polarization is vertical and antiparallel to the main magnetic
field ($y$ axis). The beam incoherence does not influence the \emph{vertical} orbital
polarization.

We predict the new effect of a \emph{radiative orbital polarization of twisted electron
beams in a magnetic field} -- \textbf{orbital Sokolov-Ternov effect}. The well-known effect is the
radiative spin polarization of electron/positron beams in storage rings caused by the
synchrotron radiation (Sokolov-Ternov effect \cite{Energy2}). It consists in the radiative
spin polarization which is acquired by unpolarized electrons and is opposite to the
direction of the main magnetic field. The reason of the effect is a dependence of
spin-flip transitions from the initial particle polarization. The standard analysis
\cite{Energy2,Landau4} can be extended on the twisted particles. It follows from the
results obtained in Refs. \cite{electpart,IvanovScattering} that quantum-electrodynamics
effects are rather similar for twisted and untwisted particles. In particular, the
amplitude of elastic scattering of two vortex electrons is well approximated by two
plane-wave
scattering amplitudes with different momentum transfers, which interfere and give
direct experimental
access to the Coulomb phase \cite{IvanovScattering}.

However, an influence of the synchrotron radiation on the orbital polarization of
twisted relativistic particles needs a detailed separate study. In this work, we
restrict ourselves to a consideration of some aspects of this problem. First of all,
we should note the evident similarity between interactions of the spin and the
intrinsic OAM with the magnetic field [see Eq. (\ref{eqHmsum})]. In particular,
energies of stationary states depend on projections of the spin and the intrinsic
OAM on the field direction. This similarity validates the existence of the effect
of the radiative orbital polarization. As well as the radiative spin polarization,
the corresponding orbital polarization acquired by unpolarized twisted electrons
should be opposite to the direction of the main magnetic field. The effect is conditioned by
transitions with a change of a projection of the intrinsic OAM. The probability of such transitions is large enough if the electron
energy is not too small. Similarly to the spin polarization, the orbital one is
observable when electrons are accelerated up to the energy of the order of 1 GeV.
The acceleration can depolarize twisted electrons but cannot vanish $L$. During the
process of the radiative polarization, the average energy of the electrons should
be kept unchanged. For additional explanations, see Supplemental Material.

Thus, a discovery of the fundamental property of the radiative orbital polarization
needs much higher energies than usual energies of twisted electron beams (about 300 keV).
We consider this as a positive factor because the twisted (vortex) states of particles
can play an important role in high-energy physics.

In this letter, we have studied the special properties of relativistic twisted
particles moving in nonuniform electric and magnetic fields. The general FW Hamiltonian
has been derived and a Stern-Gerlach-like force has been presented in terms of the
centroid parameters. Furthermore, a benchmark experiment has been proposed to
confirm the dynamics of the intrinsic OAM of twisted electrons with a very simple
experimental setup. At last,
we have predicted a new "orbital Sokolov-Ternov effect" for a twisted electron beam
in a magnetic field, which can be measured with a high-energy twisted electron beam.
The present state of quantum mechanics of twisted electrons is controversial because it uses the nonrelativistic approximation for a description of relativistic objects.
Our relativistic approach substantiates (and generalizes) the most part of previous results. We have correctly introduced the interaction of the intrinsic OAM with the electric field into quantum mechanical
equations for the first time and have corrected the equation of its motion  
in the electric field previously obtained in Ref. \cite{Bliokh2007}.
The relativistic approach shows that the Landau levels in the uniform magnetic field are not equidistant.


This work was supported by the Belarusian Republican Foundation for Fundamental Research
(Grant No. $\Phi$18D-002), by the National Natural Science Foundation of China (Grants No. 11575254 and No. 11805242),
by the National Key Research and Development Program of China (No. 2016YFE0130800),
and
by the Heisenberg-Landau program of the German Federal Ministry of Education and Research (Bundesministerium f\"{u}r Bildung und
Forschung).
A. J. S. also acknowledges hospitality and support by the Institute of Modern
Physics of the Chinese Academy of Sciences. The authors are grateful to
I. P. Ivanov for helpful exchanges.


\pagebreak

\onecolumngrid
\begin{center}
  \textbf{\large Supplemental Material to ``Relativistic quantum dynamics of twisted electron beams in arbitrary electric and magnetic fields''  }\\[.2cm]
\end{center}

\setcounter{equation}{0}
\setcounter{figure}{0}
\setcounter{table}{0}
\setcounter{page}{1}
\renewcommand{\theequation}{S\arabic{equation}}
\renewcommand{\thefigure}{S\arabic{figure}}
\renewcommand{\bibnumfmt}[1]{[S#1]}
\renewcommand{\citenumfont}[1]{#1}

In this Supplemental Material, we present some additional explanations of the \emph{orbital} Sokolov-Ternov effect (a radiative orbital polarization of twisted electron
beams in a magnetic field). We briefly consider an origin of the Sokolov-Ternov effect consisting in a radiative \emph{spin} polarization following Ref. \cite{Energy2}. Due to the latter effect, electron spins acquire the vertical polarization antiparallel to the main magnetic field. The maximum value of the spin polarization in the Sokolov-Ternov effect is $8\sqrt3/15\approx0.924$.

The effects of a radiative orbital polarization and a radiative  spin one are rather similar. As a result of radiative transitions, electrons tend to go to a state with the lowest energy. When the kinetic energy of the electrons is kept by a focusing field, this is the state with the lowest potential energy. In the vertical magnetic field $\bm B=B\bm e_z$, it is defined by
\begin{equation}
\begin{array}{c}
W=-\bm\mu\cdot\bm B,
\end{array}
\label{SuppMmu}
\end{equation}
where the operator $\bm\mu$ is given by Eq. (7). For electrons ($e=-|e|$), the predominant direction of the spins and intrinsic OAMs is antiparallel to the $z$ axis.

The probability of \emph{spin-flip} transitions has the form \cite{Energy2}
\begin{equation}
\begin{array}{c}
w=\frac{1}{2\tau}\left(1+\frac{8\sqrt3}{15}\zeta\right),
\end{array}
\label{spinflip}
\end{equation}
where the relaxation time is equal to
\begin{equation}
\begin{array}{c}
\tau=\frac{8\hbar^2}{5\sqrt3\,mce^2}\left(\frac{mc^2}{\epsilon}\right)^2\left(\frac{B_0}{B}\right)^3,
\end{array}
\label{reltime}
\end{equation} $\zeta=\pm1$ is the initial spin projection onto the $z$ axis, and $B_0=m^2c^3/(|e|\hbar)=4.41\times10^{13}$ G. Certainly, there are also radiative transitions with a conservation of the spin projection.

The radiative transitions with a change of the vertical projection of the intrinsic OAM, $L_z=\mathfrak{m}\hbar$ can be described in a similar way. The selection rule for electric dipole transitions is $\Delta\mathfrak{m}=0,\pm1$. Transitions between levels with $|\Delta\mathfrak{m}|>1$ are forbidden and can be disregarded because their probabilities are much less than probabilities of electric dipole transitions. In this case, the probability of transitions \emph{with a change of} the quantum number $\mathfrak{m}$ can be presented in the form
\begin{equation}
\begin{array}{c}
w_{\mathfrak{m}}=\frac{1}{\tau'}\left(a_{\mathfrak{m}}+\chi b_{\mathfrak{m}}\right),
\end{array}
\label{OAMchng}
\end{equation} where $\tau'$ is the relaxation time, $a_{\mathfrak{m}},\,b_{\mathfrak{m}}$ are some positive coefficients dependent on the initial quantum number $\mathfrak{m}$, and $\chi=\mathfrak{m}-\mathfrak{m}'$ is the
difference between the initial and final quantum numbers.
Transitions with a decrease of the vertical projection of the intrinsic OAM ($\chi>0$) are more probable.

A description of OAM-dependent effects, unlike spin-dependent ones, needs $2L+1$ transport equations. For the spin-flip transitions, the two transport equations are given by \cite{Energy2}
\begin{equation}
\begin{array}{c}
\frac{dn_-}{dt}=n_+w_{\zeta=1}-n_-w_{\zeta=-1},\\ \frac{dn_+}{dt}=n_-w_{\zeta=-1}-n_+w_{\zeta=1},\quad n_-+
n_+=n=const,
\end{array}
\label{spintrp}
\end{equation} where the plus and minus signs denote the states with the positive and negative spin projections. For twisted electrons, the transitions changing the quantum number $\mathfrak{m}$ are described by the following transport equations:
\begin{eqnarray}
\frac{dn_{-L}}{dt}&=&n_{-L+1}w_{-L+1,\,\chi=1}-n_{-L}w_{-L,\,\chi=-1},\nonumber \\
\frac{dn_{\mathfrak{m}}}{dt}&=&n_{\mathfrak{m}+1}w_{\mathfrak{m}+1,\,
\chi=1}+n_{\mathfrak{m}-1}w_{\mathfrak{m}-1,\,\chi=-1}  -n_{\mathfrak{m}}(w_{\mathfrak{m},\,\chi=1}+w_{\mathfrak{m},\,\chi=-1}),
~~~~ -L<\mathfrak{m}<L, \nonumber \\
\frac{dn_{L}}{dt}&=&n_{L-1}w_{L-1,\,\chi=-1}-n_{L}w_{L,\,\chi=1}, 
~~~~
n_{-L}+ n_{-L+1}+\dots+n_L=n=const.
\label{OAMtransporteq}
\end{eqnarray}

While the transport equations for the intrinsic OAM are much more cumbersome than those for the spin, the dynamics of the intrinsic OAM and the spin is similar. The result is the final orbital polarization antiparallel to the $z$ axis.

A quantitative description of the dynamics of the intrinsic OAM and the determination of the final vertical orbital polarization needs computer calculations.

\end{document}